\documentclass[review]{elsarticle}
\usepackage{hyperref}
\usepackage[margin=1in]{geometry}
\usepackage{graphicx}
\usepackage{amsmath}
\usepackage{placeins}
\usepackage{comment}

\journal{Journal of Nuclear Materials}
\begin{document}
\begin{frontmatter}
\title{Effect of strain and temperature on the threshold displacement energy in body-centered cubic iron}

\author[davisa,berka]{Benjamin Beeler\corref{qwe}}
\cortext[qwe]{Corresponding author}
\ead{bwbeeler@ucdavis.edu}
\author[berka]{Mark Asta}
\author[berkb]{Peter Hosemann}
\author[davisa,davisb]{Niels Gr$\o$nbech-Jensen}
\address[davisa]{Department of Mechanical and Aerospace Engineering, University of California, Davis, CA 95616}
\address[berka]{Department of Materials Science, University of California, Berkeley, CA 94720}
\address[berkb]{Department of Nuclear Engineering, University of California, Berkeley, CA 94720}
\address[davisb]{Department of Mathematics, University of California, Davis, CA 95616}

\begin{abstract}
The threshold displacement energy (TDE) is the minimum amount of kinetic energy required to displace an atom from its lattice site.  The magnitude of the TDE displays significant variance as a function of the crystallographic direction, system temperature and applied strain, among a variety of other factors.  It is critically important to determine an accurate value of the TDE in order to calculate the total number of displacements due to a given irradiation condition, and thus to understand the materials response to irradiation.  In this study, molecular dynamics simulations have been performed to calculate the threshold displacement energy in body-centered cubic iron as a function of strain and temperature.  With applied strain, a decrease of the TDE of up to approximately 14 eV was observed.  A temperature increase from 300 K to 500 K can result in an increase of the TDE of up to approximately 9 eV.
\end{abstract}
\end{frontmatter}


\section{Introduction}

Structural materials for nuclear fission and fusion applications experience a dynamic temperature, irradiation and mechanical loading environment.  Understanding materials response to irradiation in this dynamic environment is one of the key issues for the design and operation of nuclear power systems.  Lifetime and operational restrictions are typically influenced by limitations associated with the degradation of structural materials properties under irradiation.  The ability to accurately predict and understand behavior of structural materials under short term and long term irradiation is thus important for designing improved materials for nuclear power applications. 

Extensive computational research has been performed in order to better understand radiation damage in widely used structural materials, including ferritic alloys based on body-centered cubic (BCC) iron (Fe) \cite{bacon1993, stoller1996, gao1996, gao1997, becquart1997, stoller1999, malerba2006}.  Common areas of interest with regards to radiation damage simulations have been general damage behavior \cite{averback1998, phythian1995, bacon1994}, variations in primary knock-on atom (PKA) energy \cite{caturla2000, zinkle1993, zarkadoula2013}, variations in simulation temperature during irradiation \cite{gao1997, phythian1995, soneda1998, bacon2004, robinson2012}, the behavior and effect of extrinsic particles \cite{hayward2010, ackland2004, terentyev2006, yang2014}, and the effect of strain \cite{beeler2015, miyashiro2011, di2013}.  However, there have been fewer modeling investigations into the threshold displacement energy.  

The threshold displacement energy (TDE) is the minimum amount of kinetic energy required to displace an atom from its lattice site.  This quantity plays a key role in radiation damage theory, where it is implemented within the Kinchin-Pease \cite{kinchinpease} or the Norgett-Robinson-Torrens (NRT) \cite{norgett1975} equations, which state that the amount of damage is proportional to the ratio of the amount of deposited energy to an effective TDE.  Therefore, it is critically important to determine an accurate value of the TDE in order to calculate the total number of displacements due to a given irradiation condition, and thus understand the materials response to irradiation.

There is a long, but somewhat limited, history of investigating the TDE in BCC Fe.  Nordlund, \textit{et al.} \cite{nordlund2006} provide a comprehensive review of the early history of TDE studies and it will not be repeated here.  A sampling of studies include that of Bacon, \textit{et al.} \cite{bacon1993} who investigated BCC Fe at 0 K and calculated the TDE as a function of angle with a Finnis-Sinclair potential.  Ackland, \textit{et al.} \cite{ackland1997} performed a similar study at 0 K using a many-body Finnis-Sinclair potential to investigate the effects of dilute Cu solute additions on the TDE.  For application in engineering environments, the TDE needs to be studied at temperature and under non-zero strain conditions.  Nordlund \textit{et al.} \cite{nordlund2006} performed an extensive study of 11 separate interatomic potentials to investigate the lower bound of the TDE in BCC Fe at 36 K.  Zolnikov, \textit{et al.} \cite{zolnikov2015} studied the TDE in BCC iron and vanadium with high applied tetragonal shear and varying temperature, with an emphasis on radiation induced plastic deformation.  Other studies have been performed to investigate the TDE in other material systems, such as nickel, rutile and diamond \cite{liu2015, delgado2011,  robinson2012, robinson2012b}.  A related study on the effect of strain on formation energy in Fe was studied via first principles by Chen \textit{et al.} \cite{chen2010}. 

 A recent study of the effects of applied strain on radiation damage generation in BCC Fe \cite{beeler2015} showed that under certain strain conditions, there can be significant increases or decreases in the amount of defect accumulation from a given cascade.  This led to a brief investigation into the displacement energy for a single PKA direction in a system with and without strain.  It was shown that applied hydrostatic expansion leads to a decrease in the threshold displacement energy.  A more expansive investigation into the TDE as a function of strain and temperature is thus warranted, and is the focus of this study.

In this paper, molecular dynamics (MD) \cite{abraham1986, allen1987} simulations have been performed to calculate the TDE in BCC Fe at 300 K and 500 K, under three separate strain conditions - unstrained, applied tetragonal shear (Bain strain), and applied hydrostatic expansion - for ten individual PKA directions.  The probability of Frenkel pair production is determined as a function of PKA direction, strain and temperature.  Probability curves are averaged to determine an effective value of the TDE, above which a Frenkel pair is more than 50 $\%$ likely to form.  The minimum amount of kinetic energy required to create a Frenkel pair is also calculated as a function of strain, temperature and PKA direction.

\section{Computational Details}
Molecular dynamics simulations are performed utilizing the LAMMPS \cite{plimpton1995} software package and the Embedded-Atom Method (EAM) \cite{daw1984} interatomic potential developed for BCC Fe by Mendelev \textit{et al.} \cite{mendelev2003} splined to a Ziegler, Biersack and Littmark (ZBL) \cite{zbl} potential at small distances.  To generate curves of the probability of Frenkel pair production as a function of PKA energy (section 3.1), a BCC supercell containing 16,000 atoms is equilibrated for 100 ps at a given temperature in an NPT ensemble.  A strain is applied to the supercell and the strained lattice is allowed to equilibrate for 100 ps in an NVT ensemble.  An atom is then given extra kinetic energy, with the velocity directed in varying prescribed directions.  The time step is set to 0.2 fs and the simulation is run for 30000 steps.  We utilize the GJF thermostat \cite{gjf2013, gjf2014} (usage in LAMMPS: fix langevin gjf) due to its robust configurational sampling properties.  The damping parameter (analogous to relaxation time) is set to 1 ps.  The number of stable Frenkel pairs is determined after 6 ps via a Wigner-Seitz cell based algorithm \cite{hayward2010}.  Thus, this analysis does not take into account long-time thermal diffusion.  

PKA energies are varied over a range of approximately 100 eV, in steps of 2.5-10 eV, to properly sample the relevant energy range.  The analysis includes ten randomly oriented PKA directions that adequately sample the BCC crystal cell \cite{beeler2015}.  These ten directions are defined in Appendix A.  For each strain state and PKA direction, 64-100 independent simulations are performed, each simulation with an independent distribution of initial velocities.  

For the calculation of the minimum value of the TDE (section 3.2), a similar approach was used, employing 2000 atom cells equilibrated for 6 ps after application of strain. Since PKA energies are much lower in determining the minimum TDE, supercells are not required to be as large.  PKA energies are varied in steps of 2 eV, to properly sample the relevant energy range.  The same set of PKA directions is utilized, as outlined in Appendix A.  

The two strain states investigated are 2$\%$ hydrostatic expansion and 5$\%$ Bain strain.  Hydrostatic expansion is defined as an isotropic increase in the lattice constant for the x, y and z directions.  Tetragonal shear strain (Bain strain) is defined as elongation in the z direction, with accompanying contraction in the x and y directions to ensure constant volume.  Further details regarding the specifications of the applied strain have been outlined previously \cite{beeler2015}.

As reviewed by Nordlund \cite{nordlund2006}, several different definitions of TDE can be introduced.  In the previous literature, the primary type of TDE reported is the E$^{\textrm{l}}_{\textrm{d}}$, or the lower bound of the displacement energy, where there exists a non-zero probability of creating a defect \cite{malerba2002}.  Typically, the direct experimental measurements of the TDE measure the lowest energy where a defect signal can be detected, for example by changes in resistivity.  This is analagous to the E$^{\textrm{l}}_{\textrm{d}}$.  For comparison to experiment, adjustments for beam spreading should be included \cite{nordlund2006}.  For implementation into the Kinchin-Pease or NRT equations, an average value for the TDE needs to be utilized \cite{nordlund2006,norgett1975}.  This average value is obtained from generating probability curves (probability of generating a Frenkel pair as a function of PKA energy) for respective PKA directions.  These probability curves can then be averaged over all directions to create an angle-integrated displacement probability curve \cite{nordlund2006}.  The energy at which this averaged probability curve crosses the probability 0.5 is determined to be the median TDE (E$^{\textrm{pp}}_{\textrm{d,med}}$), where pp stands for production probability.  This value takes into account not only displacement of atoms, but allows for subsequent recombination in the time-frame of the 6 ps simulation following the PKA event.  Both the E$^{\textrm{pp}}_{\textrm{d,med}}$ and the E$^{\textrm{l}}_{\textrm{d}}$ are calculated in this work.

\section{Results}
\subsection{Median Threshold Displacement Energy}
In this section, E$^{\textrm{pp}}_{\textrm{d,med}}$ is determined through a variety of simulations.  The results in Figure 1 outline the probability of Frenkel pair production as a function of the PKA energy for each of the ten random directions investigated.  Each graph includes data points and fourth-order polynomial fits for an unstrained system, a system with applied 2$\%$ hydrostatic expansion and a system with applied 5$\%$ Bain strain.  In these graphs, the red square data points and the red dashed line denote a system without external applied strain, the blue diamond data points and the blue solid line denote a system with 2$\%$ hydrostatic expansion.  and the green triangle data points and the green dashed line denote a system with 5$\%$ Bain strain.  

\begin{figure}[hp]
   \centering
   \includegraphics[width=0.8\textwidth]{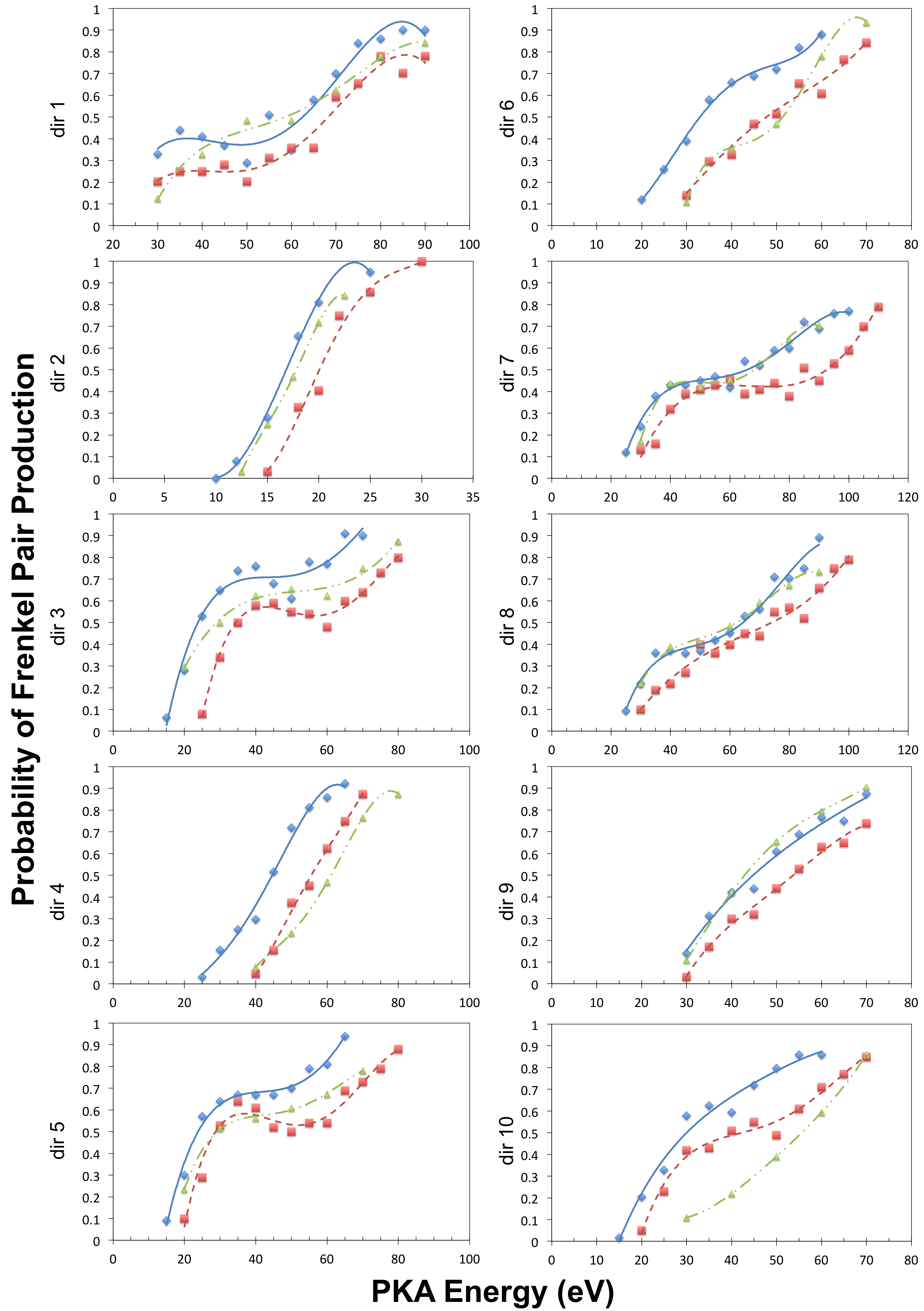} 
   \caption{Probability of Frenkel pair production as a function of PKA energy at 300 K for ten random PKA directions.  The directions are labeled from \textit{dir 1} to \textit{dir 10}.  The red square data points and the red dashed line denote a system without external applied strain.  The blue diamond data points and the blue solid line denote a system with 2$\%$ hydrostatic expansion.  The green triangle data points and green dashed line denote a system with 5$\%$ Bain strain.  Trend lines are fourth-order polynomials.}
   \label{fig:example}
\end{figure}

There are numerous interesting details from these ten graphs.  The first is the general behavior of the probability of Frenkel pair production as a function of PKA energy.  In accordance with previous results \cite{nordlund2006}, there are large ranges in energy for nearly all investigated directions where the probability to create a Frenkel pair is finite and different from one.  This suggests it would be more accurate to utilize the entire probability versus PKA energy curve to determine the actual displacements per atom under irradiation.  Also, there often exists a plateauing of the probability curve.  This can be understood as due to excess heat leading to increased recombination.  As PKA energies increase, there is locally an increase in the temperature, this can allow for increased diffusion/annihilation of defects.  However, an increase in the PKA energies also leads to more defects being produced.  These competing effects yield this plateauing behavior.  As the PKA energy increases, it finally becomes large enough to overcome this increase in recombination, leading to an increase in the probability of defect formation.  

The second key feature of the graphs is the relative probabilities for a given PKA energy with and without applied strain.  Overall, there is a clear trend that the application of 2 $\%$ hydrostatic tensile strain shifts the probability curves to lower energies, suggesting that the TDE is lower under tensile strains.  This point will be discussed in further detail below.  There is some statistical variation in the data, but there exists an overall trend of increasing probability for Frenkel pair creation in the systems with applied tensile strain.  This shows that the TDE is in fact affected by applied strain.  For comparisons between unstrained system and systems with applied Bain strain, the trend is less clear.  However, in most cases application of Bain strain results in a shift left of the probability curve, and thus a decrease in the TDE.  Thirdly, it is clear that the general behavior of the probability versus energy curves varies greatly as a function of PKA direction.  This highlights the strong crystallographic dependence of the TDE and the need for investigation over a variety of PKA directions.

Using the data in Figure 1, we want extract a value for the TDE for each direction.  The point at which the probability curve becomes greater than or equal to 0.5 is taken as the value of the TDE for that specific PKA direction.  Thus, the TDE is the energy when it becomes probable that a defect is created.  This data is tabulated in Table 1 for systems at 300 K and in Table 2 for systems at 500 K.  

\begin{table}[htbp]
\caption{The threshold displacement energy as a function of direction at 300 K for three strain conditions.  Energies given in eV.}
\begin{center}
\begin{tabular}{|c|c|c|c|}
	\hline
	& Unstrained & Hydrostatic 2$\%$ & Bain 5$\%$ \\
	 \hline
	 dir 1 & 68 & 63 & 59 \\
	 dir 2 & 20 & 17 & 18 \\
	 dir 3 & 29 & 24 & 29 \\
	 dir 4 & 56 & 45 & 61 \\
	 dir 5 & 29 & 24 & 29 \\
	 dir 6 & 48 & 33 & 51 \\
	 dir 7 & 93 & 64 & 68 \\
	 dir 8 & 74 & 64 & 62 \\
	 dir 9 & 54 & 46 & 46 \\
	 dir 10 & 42 & 30 & 56 \\
	 \hline
\end{tabular}
\end{center}
\label{default}
\end{table}%

\begin{table}[htbp]
\caption{The threshold displacement energy as a function of direction at 500 K for three strain conditions.  Energies given in eV.}
\begin{center}
\begin{tabular}{|c|c|c|c|}
	\hline
	& Unstrained & Hydrostatic 2$\%$ & Bain 5$\%$ \\
	 \hline
	 dir 1 & 81 & 54 & 78 \\
	 dir 2 & 23 & 18 & 22 \\
	 dir 3 & 68 & 30 & 60 \\
	 dir 4 & 61 & 52 & 68 \\
	 dir 5 & 65 & 49 & 39 \\
	 dir 6 & 58 & 46 & 53 \\
	 dir 7 & 89 & 77 & 67 \\
	 dir 8 & 82 & 69 & 57 \\
	 dir 9 & 60 & 51 & 42 \\
	 dir 10 & 54 & 37 & 59 \\
	 \hline
\end{tabular}
\end{center}
\label{default}
\end{table}%

\FloatBarrier

The probability curves displayed in Figure 1 are arithmetically averaged, creating a single angle-integrated probability curve for each of the three strain conditions.  The resulting three probability curves are displayed in Figure 2, for the system at 300 K and 500 K.  In order to determine the value of E$^{\textrm{pp}}_{\textrm{d,med}}$, a forrth order polynomial is fit to the data and the E$^{\textrm{pp}}_{\textrm{d,med}}$ is calculated from the point where the polynomial fit crosses 0.5 probability.  

\begin{figure}[hp]
   \centering
   \includegraphics[width=\textwidth]{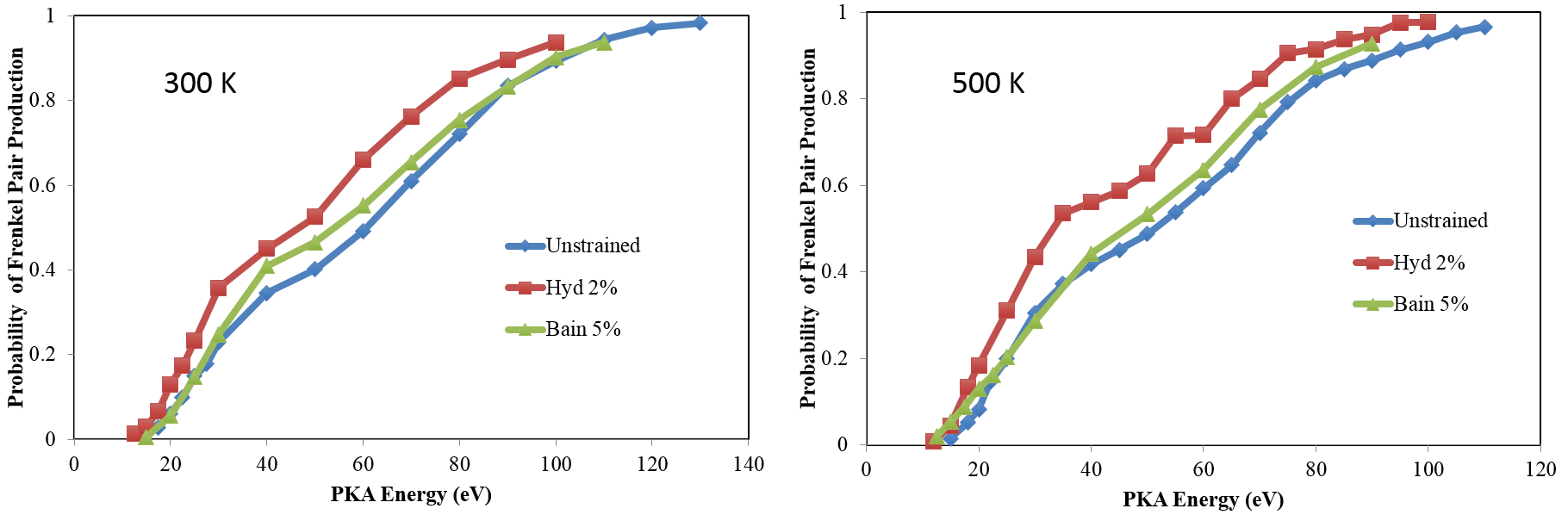} 
   \caption{The angle-integrated probability curves at 300 K and 500 K for three strain conditions.  The blue diamond data points denote a system without external applied strain.  The red square data points denote a system with 2$\%$ hydrostatic expansion. The green triangle data points denote a system with 5$\%$ Bain strain.}
   \label{fig:example}
\end{figure}

Figure 2 shows that with applied hydrostatic expansion, there is a marked shift up of the probability curve for both temperatures.  Thus, for a given PKA energy, it is more likely to create a Frenkel pair in a system under 2$\%$ hydrostatic expansion, than in an unstrained system.  This is consistent with previous work \cite{beeler2015}.  This also illustrates that there is an overall decrease of approximately 14 eV in the TDE at 300 K and a decrease of approximately 13 eV at 500 K that would be used as an input into the Kinchin-Pease or NRT radiation damage models.  For applied Bain strain at 300 K, there exists a very minor shift up of the probability curve, yielding minimal changes to the average radiation damage response of the material.  This is also consistent with previous work \cite{beeler2015}.  However, the shift is more substantial for simulations at 500 K.  Application of 5$\%$ Bain strain at 500 K yields a decrease in the E$^{\textrm{pp}}_{\textrm{d,med}}$ of 5 eV.

\FloatBarrier

The results for temperatures at 300K and 500 K from Figure 2 are tabulated in Table 3.  The values of E$^{\textrm{pp}}_{\textrm{d,med}}$ at 300 K for the unstrained, hydrostatic 2$\%$ and Bain $5\%$ are 50.0 eV, 35.8 eV and 46.2 eV, respectively.  The values of E$^{\textrm{pp}}_{\textrm{d,med}}$ at 500 K for the unstrained, hydrostatic 2$\%$ and Bain $5\%$ are 57.4 eV, 44.4 eV and 51.7 eV, respectively.  The standard value for the TDE in iron is usually taken as 40 eV \cite{was2007, ASTMstandard}.  This value is for a system at 0 K, and thus thermal fluctuations at 300 K and 500 K are dictating an increase in the TDE to a value of 50 eV and 57 eV, respectively.   Nordlund \cite{nordlund2006} calculated the E$^{\textrm{pp}}_{\textrm{d,med}}$ at 36 K in an unstrained system utilizing the Ackland ABC potential \cite{ackland1997}, resulting in a value of 44 eV.  This corresponds to our results in that an increase in temperature is yielding an increase in the TDE.  It should be noted that the effect of temperature on the TDE is much larger than the thermal energy.  

For all strain conditions, an increase in temperature yields an increase in the E$^{\textrm{pp}}_{\textrm{d,med}}$.  For both temperatures, the unstrained system exhibits the highest displacement energy and the system under 2$\%$ hydrostatic expansion exhibits the lowest displacement energy.  The largest variance in the TDE with respect to strain has a magnitude of 14.2 eV.  The largest variance in the TDE with respect to temperature is 8.6 eV.  In the Kinchin-Pease and NRT equations, the number of displacements is inversely proportional to the TDE.  If we take as an example the unstrained system of BCC Fe, using the TDE value at 300 K yields approximately 18$\%$ more displacements than if the value of TDE at 500 K is utilized, and 18$\%$ fewer displacements than if the typical value for BCC Fe of 40 eV is utilized.  This underlines the importance of the proper implementation of the appropriate E$^{\textrm{pp}}_{\textrm{d,med}}$ for a particular system.  

In order to determine the statistical significance of these results, the standard error of the mean was determined for each data point in Figure 2.  The standard error of the mean for each data point was added to that respective data point in order to create a higher bound probability curve.  Likewise a lower bound probability curve was created by subtracting the standard error of the mean for each respective data point.  The upper bound and lower bound E$^{\textrm{pp}}_{\textrm{d,med}}$ is calculated from the point where the upper bound and lower bound probability curves, respectively, cross 0.5 probability.  This can provide us with a most probable range for the values of E$^{\textrm{pp}}_{\textrm{d,med}}$.  These ranges are tabulated in Table 3.  This data shows that although there is a slight depression in the TDE due to applied 5$\%$ Bain strain, this decrease is not statistically significant given the finite set of PKA directions analyzed.  However, this data confirms that the decrease in the TDE due to applied 2$\%$ hydrostatic expansion is statistically significant.

\begin{table}[htbp]
\caption{The median threshold displacement energy in BCC Fe at three strain conditions and two temperatures.  Energies given in eV.  The range of $\pm$ one standard deviation is given in parentheses.}
\begin{center}
\begin{tabular}{|c|c|c|}
	\hline
	& 300 K & 500 K \\
	 \hline
	 Unstrained & 50.0 (45.1-54.7) & 57.4 (45.1-54.7) \\
	 Hyd 2$\%$ & 35.8 (30.2-40.9) & 44.4 (38.2-50.0) \\
	 Bain 5$\%$ & 46.2 (38.8-52.3) & 51.7 (44.0-59.3) \\
	 \hline
\end{tabular}
\end{center}
\label{default}
\end{table}

\FloatBarrier
\subsection{Lower Bound of the Threshold Displacement Energy}

Typically, the direct experimental measurements of the TDE measure the lowest incident electron energy where a defect signal can be detected, for example by changes in resistivity.  This is analogous to the lower bound of the threshold displacement energy, E$^{\textrm{l}}_{\textrm{d}}$.  The E$^{\textrm{l}}_{\textrm{d}}$ is the lowest possible value of kinetic energy such that a Frenkel pair can be created.  In this work, the minimum probability to meet the criterion for defect production is 0.01.  The results for the E$^{\textrm{l}}_{\textrm{d}}$ as a function of PKA direction and temperature are displayed in Figure 3, for all three strain conditions.  

From Figure 3, it is found that the magnitude of E$^{\textrm{l}}_{\textrm{d}}$ varies from 10 eV at the minimum to 26 eV at the maximum across all PKA directions and temperatures.  There is a strong dependence of the E$^{\textrm{l}}_{\textrm{d}}$ on crystallographic direction.  These results are numerically averaged to determine the average E$^{\textrm{l}}_{\textrm{d}}$ for each strain condition and temperature.  These results are tabulated in Table 4.

\begin{figure}[hp]
   \centering
   \includegraphics[width=\textwidth]{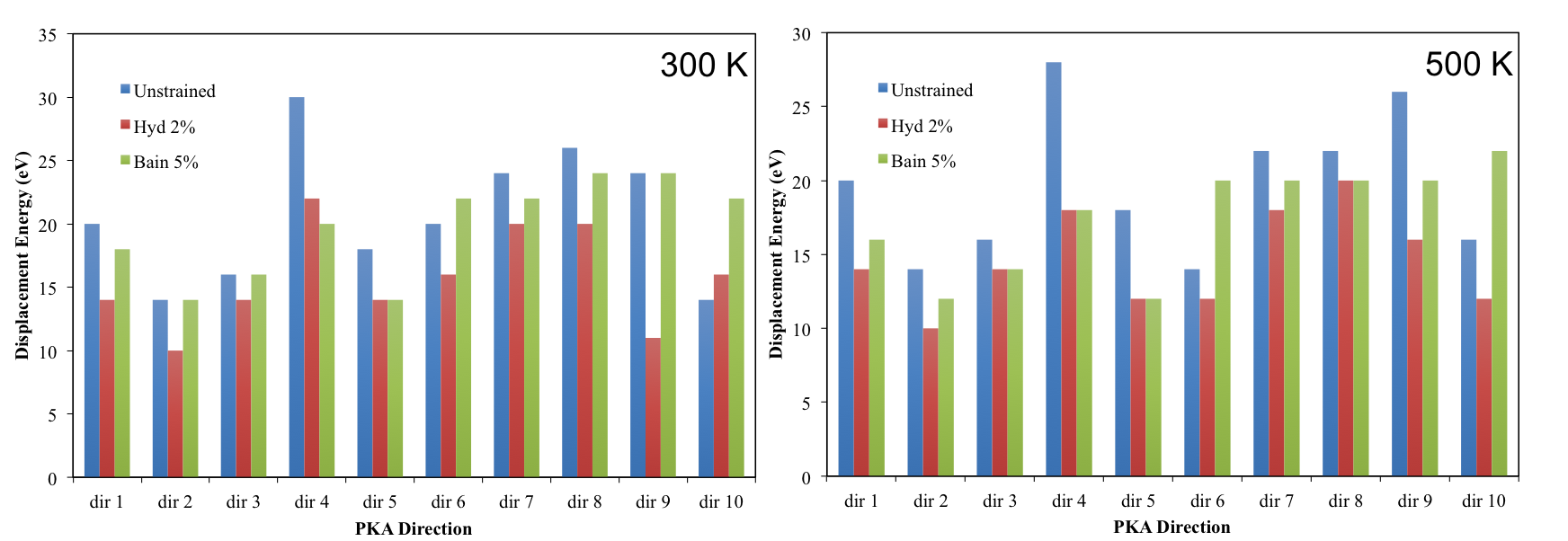} 
   \caption{The lower bound threshold displacement energy as a function of direction at 300 K and 500 K.  The blue bars denote a system without external applied strain.  The red bars denote a system with 2$\%$ hydrostatic expansion. The green bars denote a system with 5$\%$ Bain strain.}
   \label{fig:example}
\end{figure}

\FloatBarrier

\begin{table}[htbp]
\caption{The lower bound of the threshold displacement energy in BCC Fe at three strain conditions and two temperatures.  Energies given in eV.  Plus/minus is the standard error of the mean.}
\begin{center}
\begin{tabular}{|c|c|c|}
	\hline
	& 300 K & 500 K \\
	 \hline
	 Unstrained & 20.6 $\pm 1.69$ & 19.6 $\pm 1.54$ \\
	 Hyd 2$\% $  & 15.7 $\pm 1.25$ & 14.6 $\pm 1.03$ \\
	 Bain 5$\%$  & 19.6 $\pm 1.22$ & 17.4 $\pm 1.16$ \\
	 \hline
\end{tabular}
\end{center}
\label{default}
\end{table}

For both temperatures, the unstrained system exhibits the highest displacement energy and the system under 2$\%$ hydrostatic expansion exhibits the lowest displacement energy, consistent with the results in Table 3 for E$^{\textrm{pp}}_{\textrm{d,med}}$.  However, for all strain conditions, the E$^{\textrm{l}}_{\textrm{d}}$ at 300 K is higher than at 500 K.  This is in direct contrast to the results in Table 3.  Thus, the E$^{\textrm{l}}_{\textrm{d}}$ and the E$^{\textrm{pp}}_{\textrm{d,med}}$ exhibit opposite behaviors as a function of temperature from 300 K to 500 K.  This can be explained by the nature of these two different definitions of the TDE.  For E$^{\textrm{pp}}_{\textrm{d,med}}$, higher energy PKAs are utilized, inducing mini-cascades.  Additional thermal fluctuations can allow for increased recombination as the thermal spike anneals.  For E$^{\textrm{l}}_{\textrm{d}}$, increased thermal fluctuations can act to rapidly diffuse a created interstitial away from the resultant vacancy, creating a stable Frenkel pair.  Thus, an increase in temperature yields an increase in the probability for a set of PKAs to create a $\it{single}$ Frenkel pair (E$^{\textrm{l}}_{\textrm{d}}$), while yielding a decrease in the $\it{overall}$ probability of creating a Frenkel pair (E$^{\textrm{pp}}_{\textrm{d,med}}$).

The traditional value used for the E$^{\textrm{l}}_{\textrm{d}}$ in BCC Fe is 20 eV \cite{was2007}.  This work shows that the absolute minimum is 14 eV for an unstrained system, which is a significantly lower value than the traditional E$^{\textrm{l}}_{\textrm{d}}$.  However, the averaged results at 300 K and 500 K for the unstrained system are in line with this traditional value.  The maximum variance in the E$^{\textrm{l}}_{\textrm{d}}$ as a function of strain is 5 eV.  The maximum variance in the E$^{\textrm{l}}_{\textrm{d}}$ as a function of temperature is 2 eV.  

\section{Conclusions}
In this study, molecular dynamics simulations were performed on pure BCC Fe to investigate the effects of applied strain and temperature on the threshold displacement energy.  Two separate values for the threshold displacement energy are calculated: the median value of the displacement energy (E$^{\textrm{pp}}_{\textrm{d,med}}$) and the lower bound of the displacement energy (E$^{\textrm{l}}_{\textrm{d}}$).  Each threshold displacement energy is determined as a function of ten PKA directions, three strain conditions, and two temperatures.  It was determined that for E$^{\textrm{pp}}_{\textrm{d,med}}$, application of 2$\%$ hydrostatic strain results in a decrease of up to 14 eV and application of 5$\%$ Bain strain results in a decrease of up to 6 eV.  For E$^{\textrm{pp}}_{\textrm{d,med}}$, an increase in the temperature of the system from 300 K to 500 K can result in an increase of up to 9 eV.  For E$^{\textrm{l}}_{\textrm{d}}$, application of 2$\%$ hydrostatic strain results in a decrease of up to 5 eV and application of 5$\%$ Bain strain results in a decrease of up to 2 eV.  For E$^{\textrm{l}}_{\textrm{d}}$, an increase in the temperature of the system from 300 K to 500 K can result in a decrease of up to 2 eV.  This study clearly shows the importance of accounting for changes in strain environment and system temperature regarding the threshold displacement energy and the resulting radiation damage behavior.  

\section{Acknowledgement}
This work was supported by the US Department of Energy, project $\#$DE-NE0000536000.

\FloatBarrier

\appendix
\section{Set of random directions}

The possible set of PKA directions in the body-centered cubic crystal structure can be defined by two angles, $\theta$ and $\phi$, as illustrated in Figure A.4.  The set of PKA directions is chosen by utilizing a random number generator to select $\theta$ and $\phi$ values, varying from 0 to $\pi$/2 and 0 to $\pi$/4, respectively.  The specific $\theta$ and $\phi$ values are provided in Table A.5, along with the corresponding \textit{h}, \textit{k}, \textit{l} values.  These \textit{h}, \textit{k}, \textit{l} values are normalized such that \textit{h} is equal to 1.  It was found that this set of directions adequately samples the BCC crystal structure.  Tests were performed utilizing a larger subset of directions, yielding negligible variance in results.

\begin{figure}[hp]
   \centering
   \includegraphics[width=0.5\textwidth]{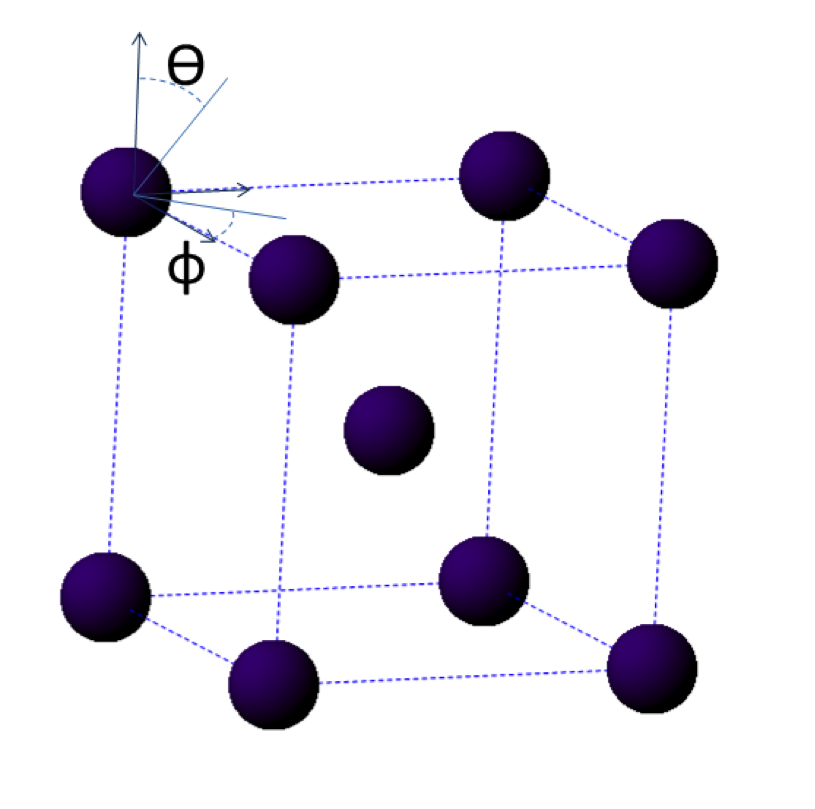} 
   \caption{Random directions were selected for primary knock-on atoms to sample the complete set of directions within a BCC crystal system.  The angle theta ($\theta$) was allowed to vary from 0 to $\pi$/2.  The angle phi ($\phi$) was allowed to vary from 0 to $\pi$/4.}
   \label{fig:example}
\end{figure}

\begin{table}[htbp]
\caption[c]{The set of ten random directions utilized in all simulations.  Values of $\theta$ and $\phi$ are given in degrees.  Values of \textit{h}, \textit{k}, \textit{l} are normalized such that \textit{h} is equal to 1.}
\begin{center}
\begin{tabular}{|c|c|c|c|c|c|}
	\hline
	& $\theta$ & $\phi$ & h & k & l \\
	 \hline
	 dir 1 & 27 & 24 & 1 & 2.25 & 4.83 \\
	 dir 2 & 17 & 5 & 1 & 11.43 & 37.53 \\
	 dir 3 & 24 & 15 & 1 & 3.73 & 8.68 \\
	 dir 4 & 31 & 11 & 1 & 5.14 & 8.72 \\
	 dir 5 & 24 & 14 & 1 & 4.01 & 9.28 \\
	 dir 6 & 86 & 26 & 1 & 2.05 & 0.16 \\
	 dir 7 & 37 & 38 & 1 & 1.28 & 2.16 \\
	 dir 8 & 35 & 44 & 1 & 1.04 & 2.06 \\
	 dir 9 & 85 & 45 & 1 & 1 & 0.12 \\
	 dir 10 & 65 & 4 & 1 & 14.30 & 6.68 \\
	 \hline
\end{tabular}
\end{center}
\label{default}
\end{table}
 
\FloatBarrier

\bibliography{bibliography_ben}

\end{document}